\documentclass[12pt]{iopart}
\usepackage{iopams}
\usepackage{graphicx}
\begin{document}

\title[Normal metal-$s$+$g$-wave superconductor tunnelling spectroscopy]
{Directional tunnelling spectroscopy of a normal metal-$s$+$g$-wave superconductor junction}

\author{P Pairor\dag and M F Smith\ddag}

\address{\dag\ School of Physics, Institute of Science,
        Suranaree University of Technology, 111 University Ave.,
        Nakhon Ratchasima, 30000 Thailand}

\address{\ddag\ Department of Physics, University of Toronto,
        60 St. George St., Toronto, Ontario, M5S 1A7 Canada}

\ead{\mailto{pairor@ccs.sut.ac.th}}

\begin{abstract}
We calculate the normal metal-$s$+$g$-wave superconductor
tunnelling spectrum for various junction orientations and for two
forms of the superconducting gap, one which allows for point nodes
and the other which allows for line nodes. For a junction oriented
with its normal parallel to the $ab$ plane of the tetragonal
superconductor, we find that the tunnelling spectrum is strongly
dependent on orientation in the plane. The spectrum contains two
peaks at energies equivalent to the magnitudes of the gap function
in the direction parallel to the interface normal and in the
direction making a $\pi/4$ angle with the normal. These two peaks
appear in both superconductors with point nodes and line nodes,
but are more prominent in the latter. For the tunnelling along the
$c$ axis, we find a sharp peak at the gap maximum in the
conductance spectrum of the superconductor with line nodes,
whereas with point nodes we find a peak occurring at the value of
the gap function along the $c$ axis. We discuss the relevance of
our result to borocarbide systems.
\end{abstract}


\submitto{\JPCM}


\maketitle

\section{Introduction}
Nonmagnetic rare-earth borocarbides, such as YNi$_2$B$_2$C and
LuNi$_2$B$_2$C, are among the materials which exhibit
unconventional superconductivity. There is strong evidence in
these materials indicating that their superconducting gap function
is highly anisotropic~\cite{boaknin,izawa02} and that there exist
low-lying excitations in the superconducting
state~\cite{izawa01,jacobs,yang,nohara99,nohara00,nohara97}. The
presence of these low energy excitations implies that the gap
function has nodes (or deep minima) on the Fermi surface. The
location of these nodes and their nature, i.e., whether they are
point nodes or line nodes is still unclear. The thermal
conductivity in a magnetic field at low
temperatures~\cite{boaknin} and the dependence of the specific
heat on the magnetic field~\cite{izawa01,nohara99} suggest the
existence of line nodes like those in the cuprates and UPt$_3$.
However, the recent measurements of the $c$-axis thermal
conductivity in a rotational magnetic field along the $ab$ plane
were interpreted as evidence for a gap function with point nodes
along [100] and [010] directions~\cite{izawa02}. Further studies
which employ different experimental techniques may be necessary to
determine the detailed structure of the superconducting gap
function in momentum space. Here we suggest that directional
tunnelling spectroscopy may be useful for this purpose.

In general, the tunnelling conductance spectrum of an anisotropic
superconductor is strongly dependent on the crystal orientation
with respect to the interface plane. In the case of a $d$-wave
superconductor with vertical line nodes, it has been shown that
features in the conductance spectrum occur at voltages which
depend strongly on crystal orientation. These voltages correspond
to values of the gap function at particular points on the Fermi
surface~\cite{pairor02}. For instance, in the case of an $ab$
plane tunnelling junction, if the surface orientation of the
superconductor is not [100] or [010], the conductance spectrum
contains a peak at the voltage corresponding to the value of the
gap function in the direction parallel to the surface normal. The
observation of this feature would, in principle, allow the
directional normal metal-superconductor (NS) tunnelling
spectroscopy to map out the magnitude of the superconducting gap
function.

In this paper, we calculate the NS tunnelling spectra of
anisotropic $s$-wave superconductors with two forms of gap
functions, which have been suggested as possible candidates for
the gap in borocarbides~\cite{maki,choi}. The candidates are
allowed by the symmetry of the borocarbide crystal structure and
consistent with experimental results. It is apparent that the gap
function has four-fold symmetry and does not change sign but there
is evidence for deep minima along [100] and [010]
directions~\cite{boaknin,izawa02,izawa01,jacobs,yang,nohara99,nohara00,nohara97}.
The simplest suitable candidates are of the $s+g$-wave form. In
our calculation, we find, similar to a $d$-wave case, that
features appear in the spectra at voltages corresponding to values
of the gap at particular points on the Fermi surface depending on
surface orientation. Some of these features are characteristic of
the general $s+g$-wave gap function and are thus common to both
candidates, while one particular feature occurs only for the
candidate with point nodes. Thus, if these features can be
observed experimentally, it may be possible to determine which
form of the gap, if either, correctly describes that of the
borocarbides.

The two candidate gap functions are both $s+g$-wave and are given
by equation~(\ref{line-gap}) and (\ref{point-gap}) respectively.
\numparts
\begin{eqnarray}
&&\Delta_{k,1} = \Delta_s - \Delta_g\cos4\phi
\label{line-gap}\\
&&\Delta_{k,2} = \Delta_s -
\Delta_g\sin^2\theta\cos4\phi\label{point-gap}
\end{eqnarray} \endnumparts
where $\theta$ and $\phi$ are the polar angles in spherical
coordinates, and $\Delta_s$ and $\Delta_g$ are the $s$ and $g$
components of the gap function respectively. For an elliptical
Fermi surface, the case of line (point) nodes occurs when
$\Delta_s=\Delta_g$ in equation~(\ref{line-gap})
(equation~(\ref{point-gap})). We study both $ab$ plane and
$c$-axis tunnelling spectra for both forms of the gap function by
using the so-called Blonder-Tinkham-Klapwijk (BTK) scattering
formalism~\cite{btk} and a continuous model to describe the
electronic structure of the normal metal and superconductor.
Although detailed features of the conductance depend on the shape
of the Fermi surface, it is sufficient to use the continuous model
which gives an elliptical Fermi surface to study the positions of
the main features in the conductance spectra~\cite{pairor02}.

In the case of tunnelling into the $ab$ plane of a $s+g$-wave
superconductor, we show in this paper that there are three main
features occurring at voltages associated with the difference of
the $s$ and $g$ component of the gap function and the values of
the gap function in two particular directions with respect to the
interface normal, one in the direction parallel to the surface
normal and the other in the direction making a $\pi/4$ angle with
the surface normal. These two features are more prominent for the
$s+g$-wave superconductor with line nodes than for the
superconductor with point nodes. In the case of $c$-axis
tunnelling junctions, there is one prominent feature at the
maximum gap in the conductance spectrum of the superconductor with
line nodes, whereas for the superconductor with point nodes the
conductance spectrum contains two features, one at the maximum gap
and the other at the value of the gap function along the $c$ axis.

In Section \ref{method}, we describe the calculation method and
the assumptions used throughout this paper. Then, we provide the
detailed results and discussion of all the cases of interest at
zero temperature in Section \ref{result}. In addition to the cases
in which the nodes exist, we also consider cases in which the
superconducting gap has small but nonzero gap minima. We show all
the results in both the Andreev limit (low barrier) and the
tunnelling limit (high barrier).

\section{Assumptions and method of calculation} \label{method}
\begin{figure}
\begin{center}
\includegraphics[scale=0.7]{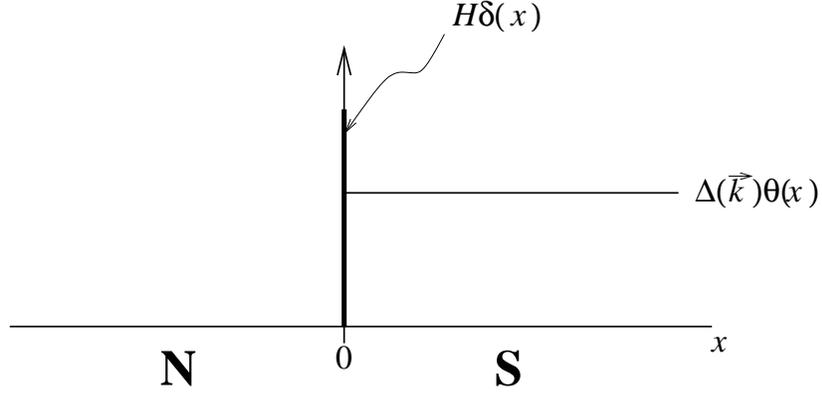}
\end{center}
\caption{\label{ns-geo}The normal metal-superconductor (NS)
junction is represented by an infinite system as shown in this
picture. The normal metal fills the $x<0$ region and the
superconductor fills the $x>0$ region. The insulator layer is
represented by a delta function of height $H$ in units of energy
per length. The gap function is taken to be zero in the normal
metal and to be nonzero and independent of $x$ in the
superconductor.}
\end{figure}

As in reference~\cite{btk}, we represent the NS junction with an
infinite system, the left half of which is a normal metal and the
right half of which is a superconductor (see figure~\ref{ns-geo}).
The insulating barrier is represented by a delta function
potential with strength $H$.
For the $ab$ plane tunnelling junctions, the interface normal
vector lies somewhere in the $ab$ plane, and for the $c$-axis
tunnelling junction the interface normal is parallel to the $c$
axis.

We take the normal metal to be cubic, and take the superconductor
to be tetragonal to describe the crystal structure of the
borocarbides. In our calculation, we ignore both the suppression
of the gap function near the NS interface and the proximity effect
for simplicity. The gap function $\Delta_k$ is taken to be as in
either equation (\ref{line-gap}) or (\ref{point-gap}).

\begin{figure}
\begin{center}
\includegraphics[scale=0.6]{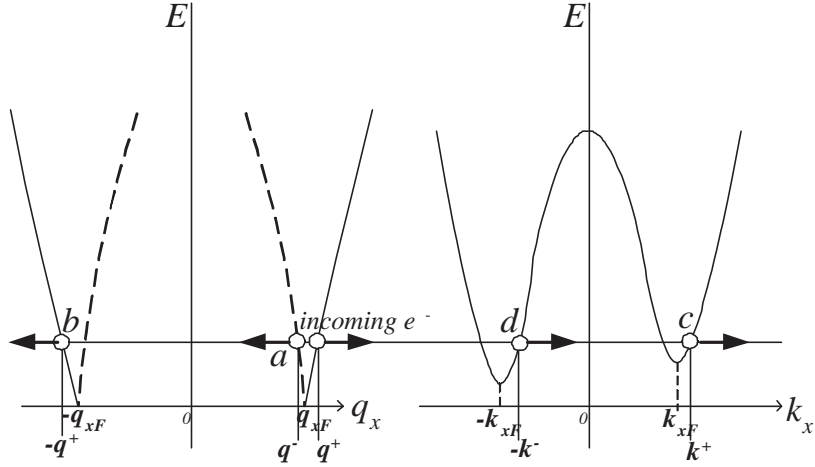}
\end{center}
\caption{\label{e-of-k}The plots of two excitation energies as a
function of $q_x$ or $k_x$, the component parallel to the
interface normal, at a particular value of $\vec{k}_\parallel$,
which is $(k_y,k_z)$. The plot on the left is for the excitation
energy of the normal metal and the plot on the right is for the
excitation energy of the superconductor. At the same energy, there
can be 4 propagating excitations for each side. However, for an
incoming electron from the normal side, the wave function of the
normal metal is a linear combination of only 3 excitations
represented by the open circles, and the wave function of the
superconductor is the sum of 2 outgoing excitations also
represented by the open circles. }
\end{figure}

The Bogoliubov-de Gennes equations that describe the excitations
of the system are
\begin{equation}
\left[
\begin{array}{cc}
\hat{O_p}+H\delta(x)-\mu & \Delta_k\Theta(x)\\
\Delta_k\Theta(x) & -\hat{O_p}-H\delta(x)+\mu
\end{array}
\right]U(\vec{r})= EU(\vec{r}) \label{bdg}
\end{equation}
where $\mu$ is the chemical potential, $\Theta(x)$ is the
Heaviside step function,
\begin{eqnarray}
\hat{O_p}=-\frac{\hbar^2}{2}\left(\frac{1}{m_{xy}}\left(\frac{\partial^2}{\partial
x^2}+\frac{\partial^2}{\partial y^2}\right)
+\frac{1}{m_z}\frac{\partial^2}{\partial z^2}\right),\nonumber
\end{eqnarray}
\begin{eqnarray} m_{xy} &=& \left\{
\begin{array}{ll}
m \mbox{ (effective mass in the normal metal), $x<0$}\\
m_{ab} \mbox{($ab$ plane effective mass in the superconductor),
$x>0$}
\end{array}
\right. \nonumber
\\ m_z &=& \left\{
\begin{array}{ll}
m\mbox{, $x<0$} \\
m_c\mbox{ ($c$-axis effective mass in the superconductor)} \mbox{,
$x>0$}
\end{array}
\right. \nonumber
\end{eqnarray}
and $U(\vec{r})$ is a two-component function:
\begin{equation}\label{two-comp-fn}
U(\vec{r}) =\left[
\begin{array}{c} u(\vec{r}) \\v(\vec{r})
\end{array}
\right]=\left[
\begin{array}{c} u_k \\ v_k
\end{array}
\right]e^{i\vec{k}\cdot\vec{r}}.
\end{equation}
After substituting $U(\vec{r})$ from equation (\ref{two-comp-fn})
into equation(\ref{bdg}), we obtain the bulk excitation energies
for the normal metal and superconductor respectively as \numparts
\begin{eqnarray}
E(\vec{q})= \pm\xi_q =
\pm\left(\frac{\hbar^2}{2m}(q_x^2+q_y^2+q_z^2)-\mu\right)\label{en}\\
E(\vec{k})= \sqrt{\xi_k^2+\Delta_k^2}\label{es}
\end{eqnarray}
\endnumparts
where for the normal metal (equation (\ref{en}))the plus and minus
signs are for electron and hole excitations respectively, and for
the superconductor (equation (\ref{es} ))
\begin{equation}
\xi_k=\frac{\hbar^2}{2}\left(\frac{k_x^2+k_y^2}{m_{ab}}
+\frac{k_z^2}{m_c}\right)-\mu.
\end{equation}
Figure~\ref{e-of-k} shows a plot of the excitation energy of the
normal metal (superconductor) as a function of $q_x$ ($k_x$), the
component along the interface normal, at a particular
$\vec{q_\parallel}=\vec{k_\parallel}=(k_y,k_z)$, the component
perpendicular to the interface normal.

The amplitudes of the excitations, $u_k$ and $v_k$, of the normal
metal are
\begin{equation}
\left[
\begin{array}{c} u_k \\v_k \end{array}
\right]=\left\{
\begin{array}{ll} \left[
    \begin{array}{c} 1 \\0 \end{array}
    \right]&\mbox{for electrons} \\ & \\
    \left[ \begin{array}{c} 0 \\1\end{array}\right]
     & \mbox{for holes}
\end{array}
\right.
\end{equation}
whereas those of the superconductor are
\begin{equation}
\left[\begin{array}{c} u_k \\v_k
\end{array}
\right]=\frac{1}{\sqrt{|E+\xi_k|^2+\Delta_k}}\left[
\begin{array}{c} \Delta_k \\ E+\xi_k
\end{array}
\right].
\end{equation}
The wave function of each side is a linear combination of all the
appropriate excitations of the same energy and the momentum that
has the same component perpendicular to the interface normal. For
$ab$ plane tunnelling, the wave functions of both sides are
therefore \numparts
\begin{eqnarray}
U_N(\vec{r})&=&\left( \left[ \begin{array}{c}
1\\0\end{array}\right]e^{iq^+x}+ a\left[ \begin{array}{c}
0\\1\end{array}\right]e^{iq^-x}+ b\left[
\begin{array}{c} 1\\0\end{array}\right]e^{-iq^+x}\right)
e^{ik_yy+ik_zz}\\
U_S(\vec{r})&=&\left( c\left[ \begin{array}{c}
u_{k+}\\v_{k+}\end{array}\right]e^{ik^+x}+d\left[
\begin{array}{c} u_{-k-}\\v_{-k-}\end{array}\right]e^{-ik^-x}\right)
e^{ik_yy+ik_zz}
\end{eqnarray}
\endnumparts
where $q^\pm$ and $k^\pm$ (see figure \ref{e-of-k}) satisfy
\numparts
\begin{eqnarray}
\hbar q^\pm &=& \sqrt{2m(\mu\pm E)-\hbar^2 k_y^2-\hbar^2 k_z^2}\\
\frac{\hbar k^\pm}{\sqrt{m_{ab}}} &=&
\sqrt{2(\mu\pm\sqrt{E^2-\Delta_k^2})-\frac{\hbar^2
k_y^2}{m_{ab}}-\frac{\hbar^2 k_z^2}{m_c}}.
\end{eqnarray}
\endnumparts
and $a$, $b$, $c$, and $d$ are the Andreev reflection, the normal
reflection, the same-branched transmission, and the cross-branched
transmission amplitudes respectively. For the $c$-axis tunnelling,
we can obtain the wave function of each side in a similar way.

Normally, the range of the energy $E$ relevant to the NS
tunnelling experiments is of order meV whereas the Fermi energy is
of order eV. Therefore, we use the following approximation for
$q^\pm$ and $k^\pm$: \numparts
\begin{eqnarray}
q^+=q^- &=& \sqrt{\frac{2m\mu}{\hbar^2}-k_y^2-k_z^2}
\nonumber\\&=& q_F\sin\theta_N\cos\phi_N\\
k^+= k^- &=& k_F\sqrt{\frac{m_{ab}}{m^*}}\sin\theta_S\cos\phi_S
\end{eqnarray}
\endnumparts
where $q_F$ is the magnitude of the Fermi wave vector of the
normal metal, and $k_F$, $m^*$ are the parameters that satisfy
$\hbar^2 k_F^2/(2m^*)=E_{F,S}$, the Fermi energy of the
superconductor. Using the conservation of the momentum parallel to
the surface, we have the following relationship between the polar
angles in spherical coordinates:
\numparts
\begin{eqnarray}
q_F\sin\theta_N\sin\phi_N&=&
k_F\sqrt{\frac{m_{ab}}{m^*}}\sin\theta_S\sin\phi_S
\label{theta-phi1}\\
q_F\cos\theta_N&=&
k_F\sqrt{\frac{m_c}{m^*}}\cos\theta_S.\label{theta-phi2}
\end{eqnarray}
\endnumparts

We obtain all the amplitudes $a$, $b$, $c$, and $d$ by applying
the following matching conditions at the interface:
\numparts
\begin{eqnarray}
U_N(x=0)=U_S(x=0)\equiv U_0\\
ZU_0=\frac{1+m/m_{ab}}{4k_F}\left(\left.\frac{\partial
U_S}{\partial x} \right|_{x=0^+} - \left.\frac{\partial
U_N}{\partial x}\right|_{x=0^-}\right)
\end{eqnarray}
\endnumparts
where $Z=mH/(\hbar^2k_F)$. Note that both matching conditions are
for the $ab$ plane tunnelling junction. For the $c$-axis
tunnelling junction, the first condition remains the same, but we
have to replace $m_{ab}$ with $m_c$ in the second condition.

In the BTK formalism, the Andreev reflection, the normal
reflection, and the two transmission probabilities are used to
obtain the current across the junction. All the reflection and
transmission probabilities are obtained from
\numparts
\begin{eqnarray}
    A&=&|a|^2\left(\frac{q^-}{q^+}\right)\\
    B&=&|b|^2\\
    C&=&|c|^2\left(|u_{k^+}|^2-|v_{k^+}|^2\right)\left(\frac{k^+}{q^+}\right)\\
    D&=&|d|^2\left(|u_{-k^-}|^2-|v_{-k^-}|^2\right)\left(\frac{k^-}{q^+}\right)
\end{eqnarray}
\endnumparts
and satisfy: $A+B+C+D=1$, i.e., the number of particles is
conserved.

On the normal metal side, we find the current across the junction
as a function of an applied voltage is
\begin{equation}
I_{NS}(V)=\frac{e\Omega}{(2\pi)^3}\int d\vec{q}\,\, v_{q_x}
[1+A(\vec{q})-B(\vec{q})][f(E_q-eV)-f(E_q)]
\end{equation}
where $\Omega$ is volume, $v_{q_x}$ is the $x$ component of the
group velocity of the incoming electron, and $f(E)$ is the
Fermi-Dirac distribution function. The conductance of the $ab$
plane junction at zero temperature is
\begin{eqnarray}
\fl G_{NS}^{ab}(V)&=&\frac{dI^{ab}_{NS}}{dV}\nonumber\\
\fl &=&\frac{me^3\Omega V}{4\pi^2\hbar^2}\int d\phi_N\int
d\theta_N
\sin^2\theta_N\cos\phi_N[1+A(V,\phi_N,\theta_N)-B(V,\phi_N,\theta_N)]
\end{eqnarray}
Similarly, the conductance of the $c$-axis tunnelling junction is
\begin{equation}
\fl G_{NS}^c(V)=\frac{me^3\Omega V}{4\pi^2\hbar^2}\int d\phi_N\int
d\theta_N\sin\theta_N\cos\theta_N[1+A(V,\phi_N,\theta_N)-B(V,\phi_N,\theta_N)]
\end{equation}
The limits of both integrals can be found by considering
equation(\ref{theta-phi1}) and (\ref{theta-phi2}).

\section{Results and discussion}\label{result}
We plot the normalized conductance as a function of applied
voltage of both $ab$ plane and $c$-axis tunnelling. We define the
normalized conductance as the conductance of the junction
normalized by its value at a high voltage, i.e., eV
$\gg\Delta_{max}$, the maximum magnitude of the gap function.
Using both forms of the gap function in equation(\ref{line-gap})
and (\ref{point-gap}), we consider 3 cases: (1) $\Delta_s =
\Delta_g$, (2) $\Delta_s = 0.9\Delta_g$, and (3) $\Delta_s =
1.1\Delta_g$. These choices of the parameter $\Delta_s/\Delta_g$
span the range allowed by the results of thermal conductivity
measurements in borocarbides which indicate that the ratio of the
gap maximum to the gap minimum is at least ten~\cite{boaknin}. All
the results are obtain for zero temperature.

\subsection{$ab$ plane tunnelling}
\begin{figure}
\begin{center}
\includegraphics[scale=0.7]{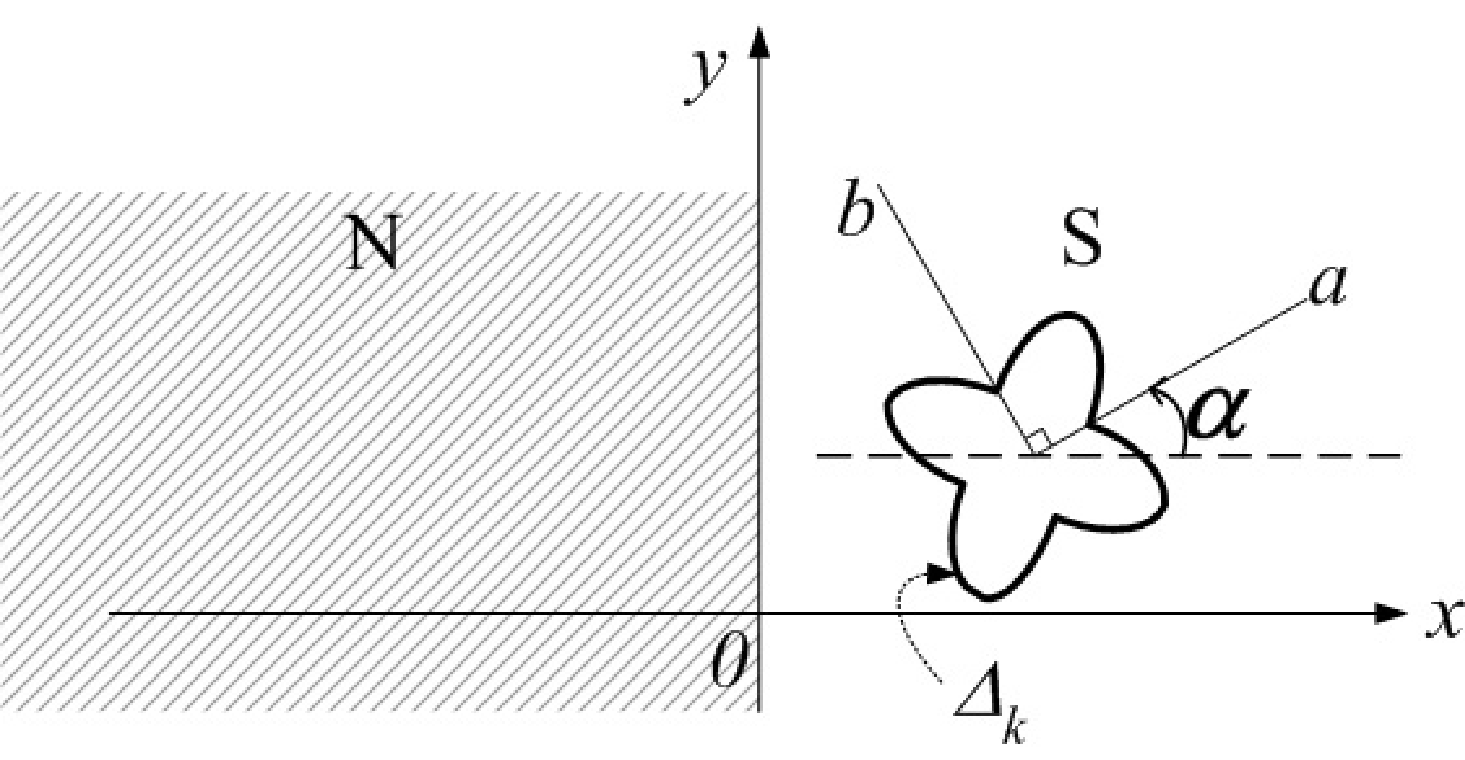}
\end{center}
\caption{\label{ab-junction}The geometry of the $ab$ plane
tunnelling NS junction. The angle between the $x$ axis and the $a$
axis, $\alpha$, defines the orientation of the junction.}
\end{figure}

Figure~\ref{ab-junction} shows the diagram of the junction that
has the interface normal on the $ab$ plane of the superconductor.
We specify the orientation in the $ab$ plane with $\alpha$, the
angle between the interface normal and the $a$ axis of the
superconductor. The gap function is a function of $\alpha$,
\numparts
\begin{eqnarray}
&&\Delta_{k^\pm,1} = \Delta_s - \Delta_g\cos4(\phi_S\mp\alpha)
\label{point-gap2}\\
&&\Delta_{k^\pm,2} = \Delta_s -
\Delta_g\sin^2\theta_S\cos4(\phi_S\mp\alpha)\label{line-gap2}
\end{eqnarray}
\endnumparts
\begin{figure}
\begin{center}
{\small(a) $\Delta_s=\Delta_g$}
\includegraphics[scale=0.4,angle=270]{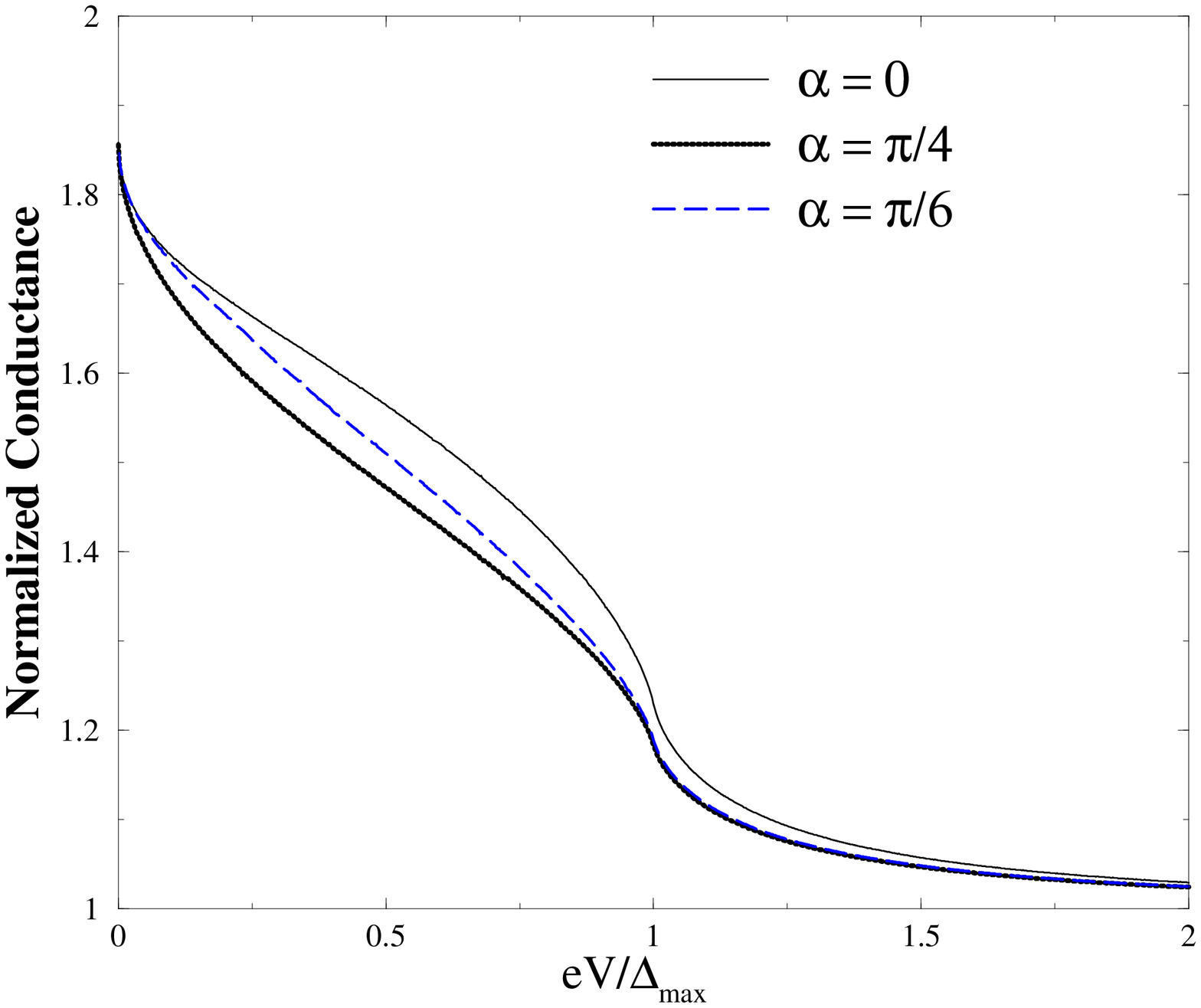}\\
{\small(b) $\Delta_s=0.9\Delta_g$}
\includegraphics[scale=0.4,angle=270]{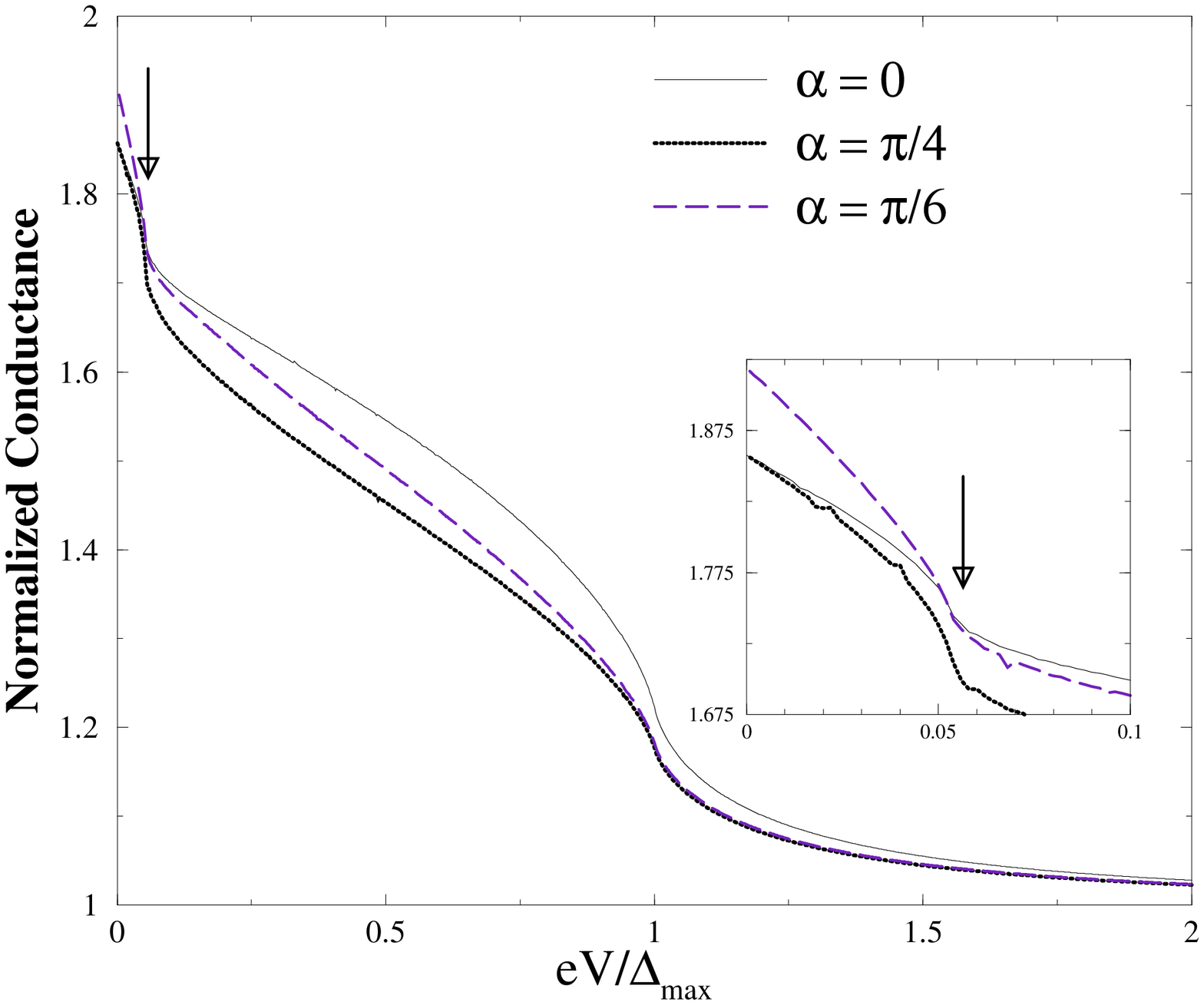}\\
{\small(c) $\Delta_s=1.1\Delta_g$}
\includegraphics[scale=0.4,angle=270]{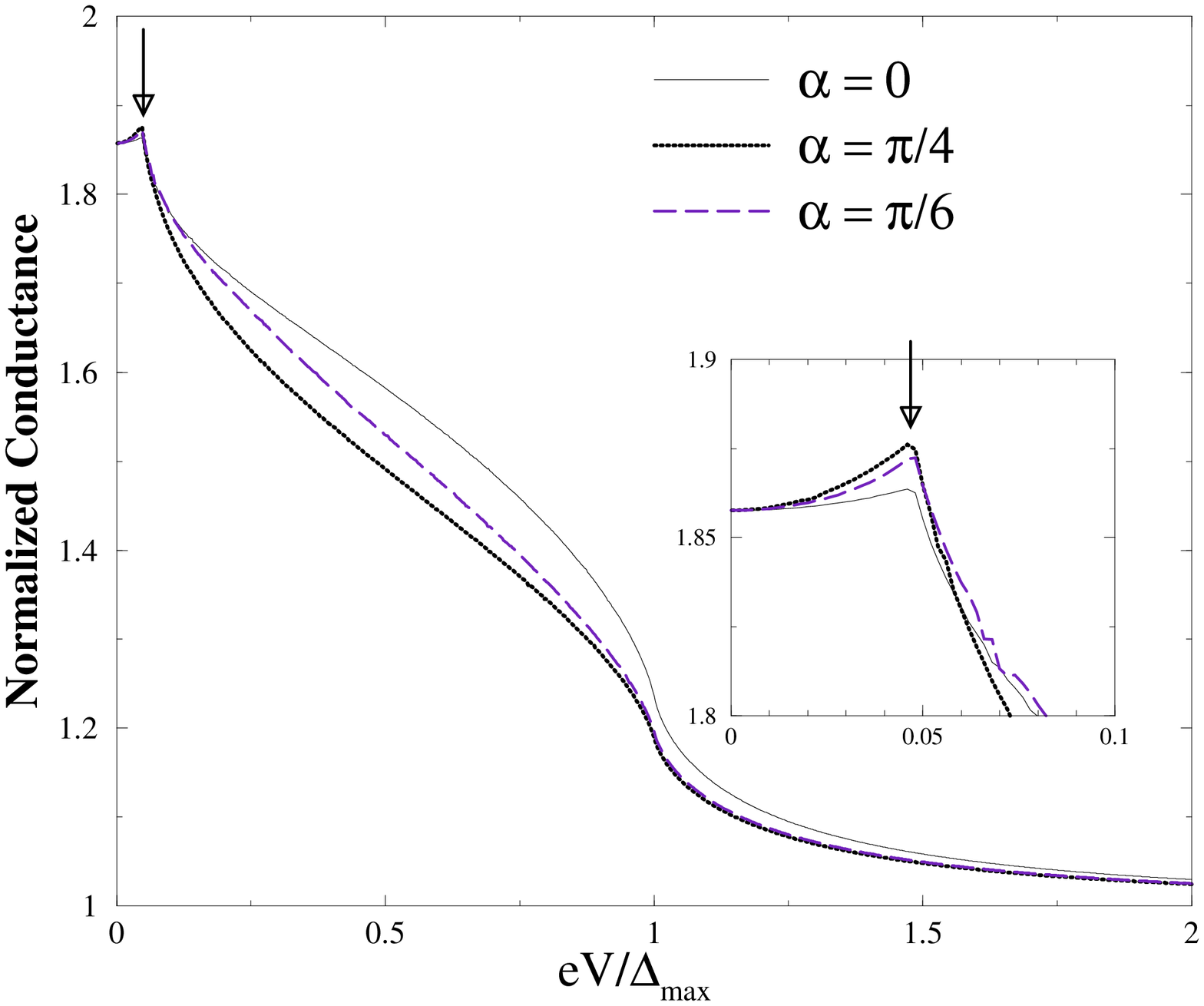}
\end{center}
\caption{\label{and-abl-tunneling}$ab$ {\it plane tunnelling
junctions.} The plots of the normalized conductance spectra when
$Z=0$ for 3 surface orientations in 3 cases of the gap function
$\Delta_{k^\pm,1} = \Delta_s - \Delta_g\cos4(\phi_S\mp\alpha)$:
(a) $\Delta_s = \Delta_g$, (b) $\Delta_s = 0.9\Delta_g$, and (c)
$\Delta_s = 1.1\Delta_g$. The arrows in (b) and (c) indicate the
feature occurring at voltage corresponding to
$|\Delta_s-\Delta_g|$. The insets in (b) and (c) are the
enlargements of the conductance plots near zero voltage.}
\end{figure}
\begin{figure}
\begin{center}
{\small(a) $\Delta_s=\Delta_g$}
\includegraphics[scale=0.4,angle=270]{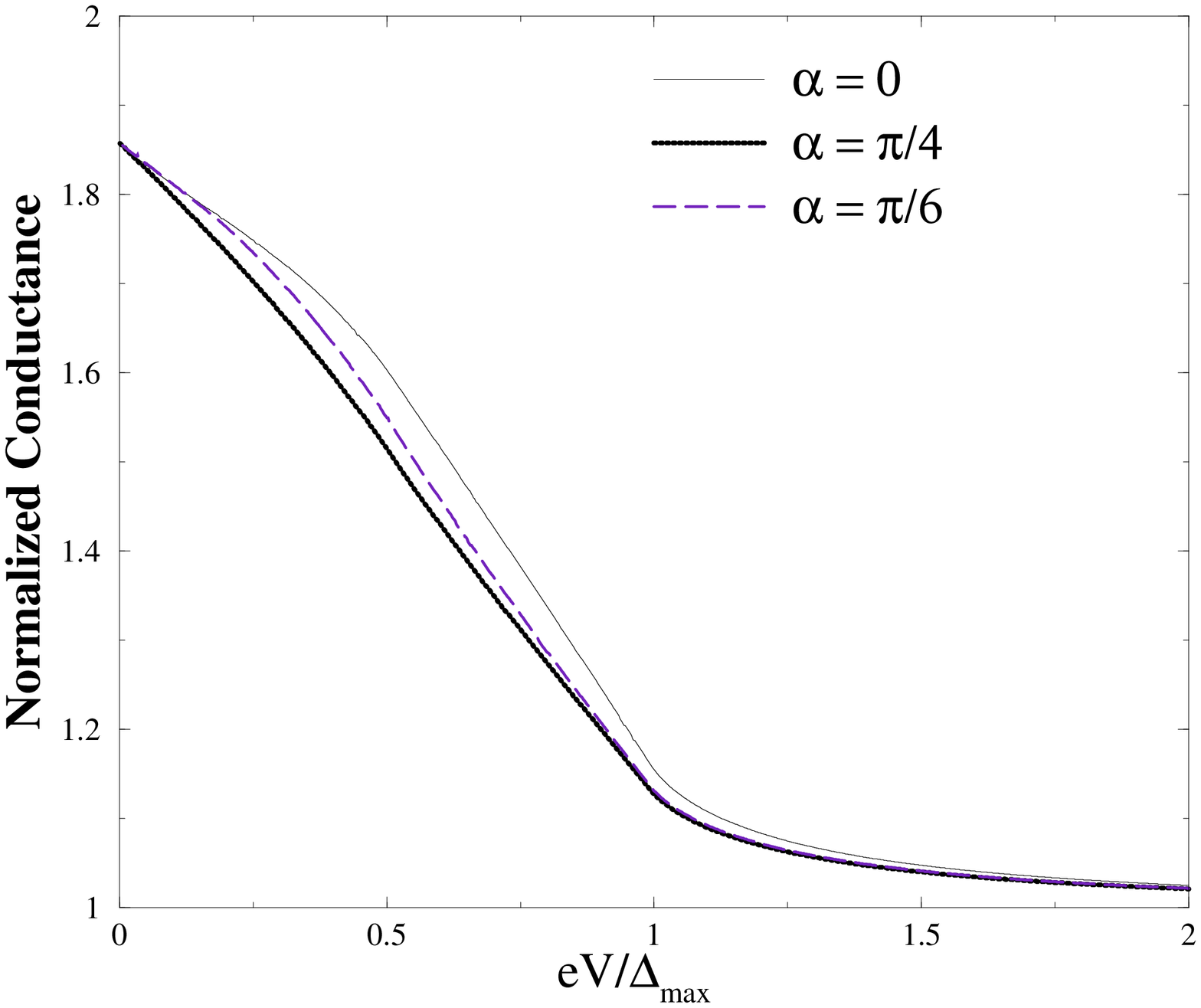}\\
{\small(b) $\Delta_s=0.9\Delta_g$}
\includegraphics[scale=0.4,angle=270]{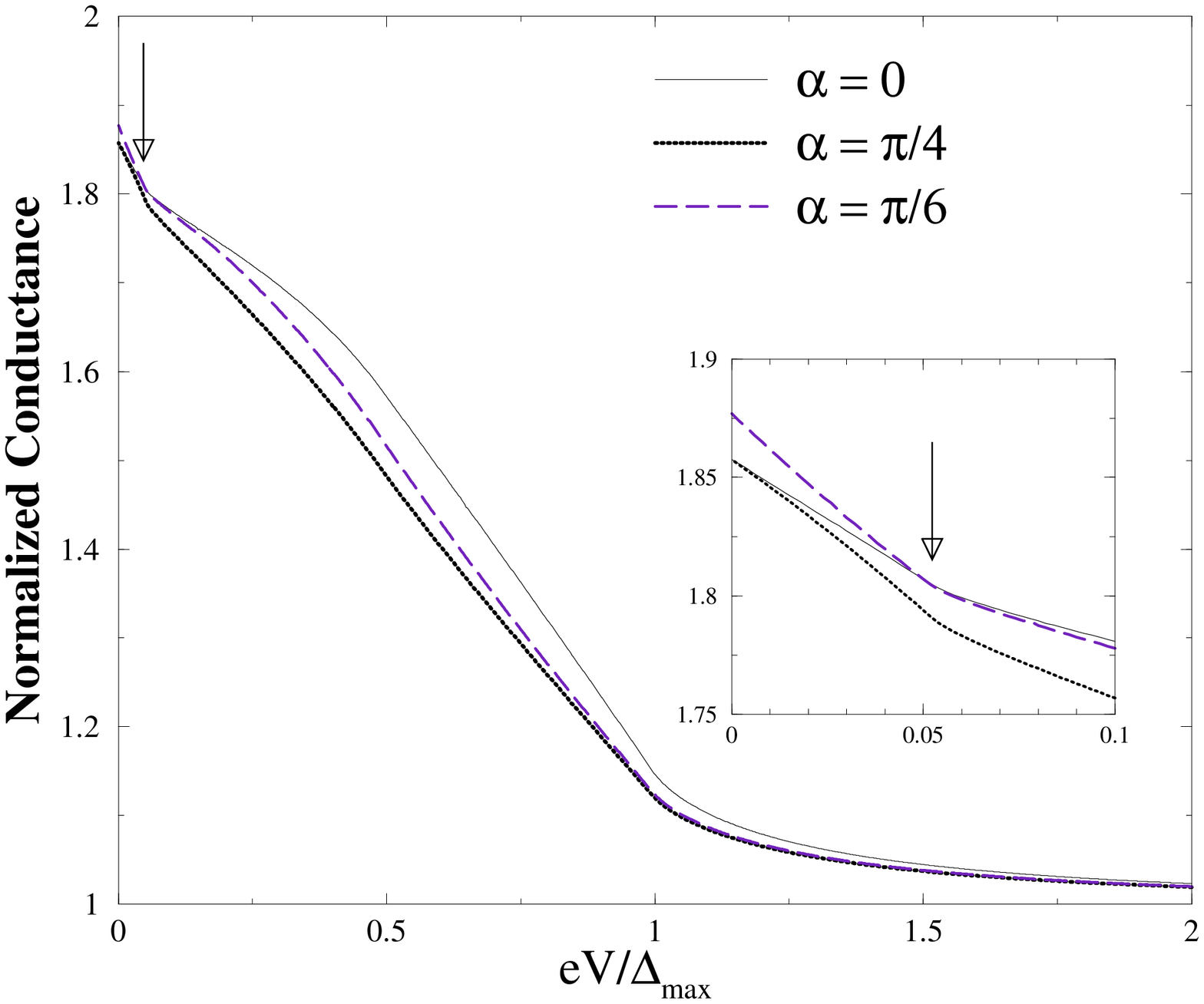}\\
{\small(c) $\Delta_s=1.1\Delta_g$}
\includegraphics[scale=0.4,angle=270]{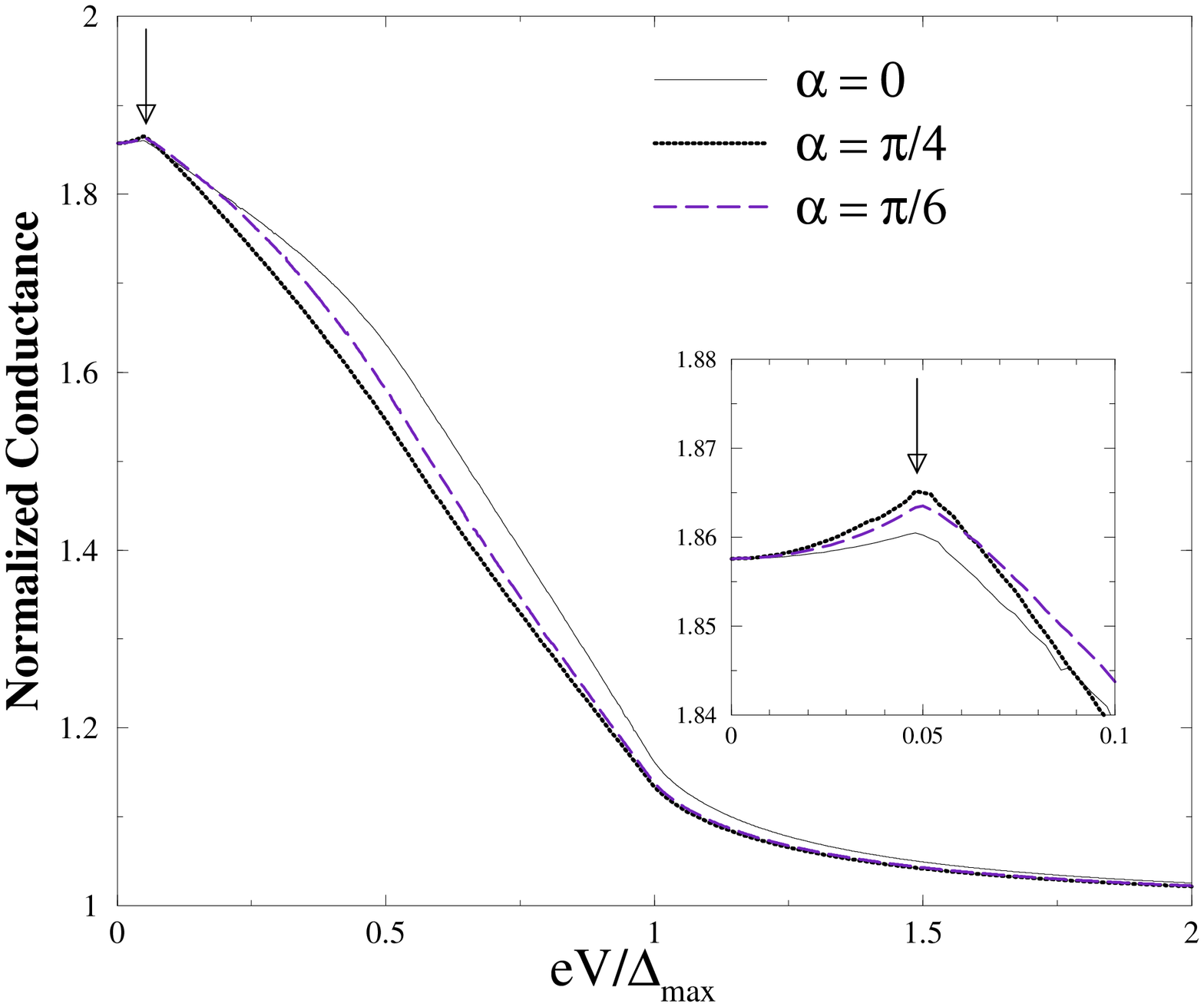}
\end{center}
\caption{\label{and-abp-tunneling}$ab$ {\it plane tunnelling
junctions.} The plots of the normalized conductance spectra when
$Z=0$ for 3 surface orientations in 3 cases of the gap function
$\Delta_{k^\pm,2} = \Delta_s -
\Delta_g\sin^2\theta_S\cos4(\phi_S\mp\alpha)$: (a) $\Delta_s =
\Delta_g$, (b) $\Delta_s = 0.9\Delta_g$, and (c) $\Delta_s =
1.1\Delta_g$. The arrows in (b) and (c) indicate the feature
occurring at voltage corresponding to $|\Delta_s-\Delta_g|$. The
insets in (b) and (c) are the enlargements of the conductance
plots near zero voltage.}
\end{figure}

In the Andreev limit (small $Z$), the conductance spectrum depends
very little on the junction orientation. As shown in
figure~\ref{and-abl-tunneling} for $\Delta_{k,1}$ and in figure
\ref{and-abp-tunneling} for $\Delta_{k,2}$, the conductance
spectra in all cases have the inverted gap structure. Note that in
the case of $\Delta_{k,2}$ in figure~\ref{and-abp-tunneling},
there is a slight kink at the voltage associated with the value of
the gap function along the $c$ axis (eV = $\Delta_s$). Also note
that when $\Delta_s\neq\Delta_g$, there is a feature occurring at
eV $=|\Delta_s-\Delta_g|$ as marked by the arrows in figure
\ref{and-abl-tunneling} (b),(c) and figure \ref{and-abp-tunneling}
(b),(c).

\begin{figure}
\begin{center}
{\small(a) $\Delta_s=\Delta_g$}
\includegraphics[scale=0.4,angle=270]{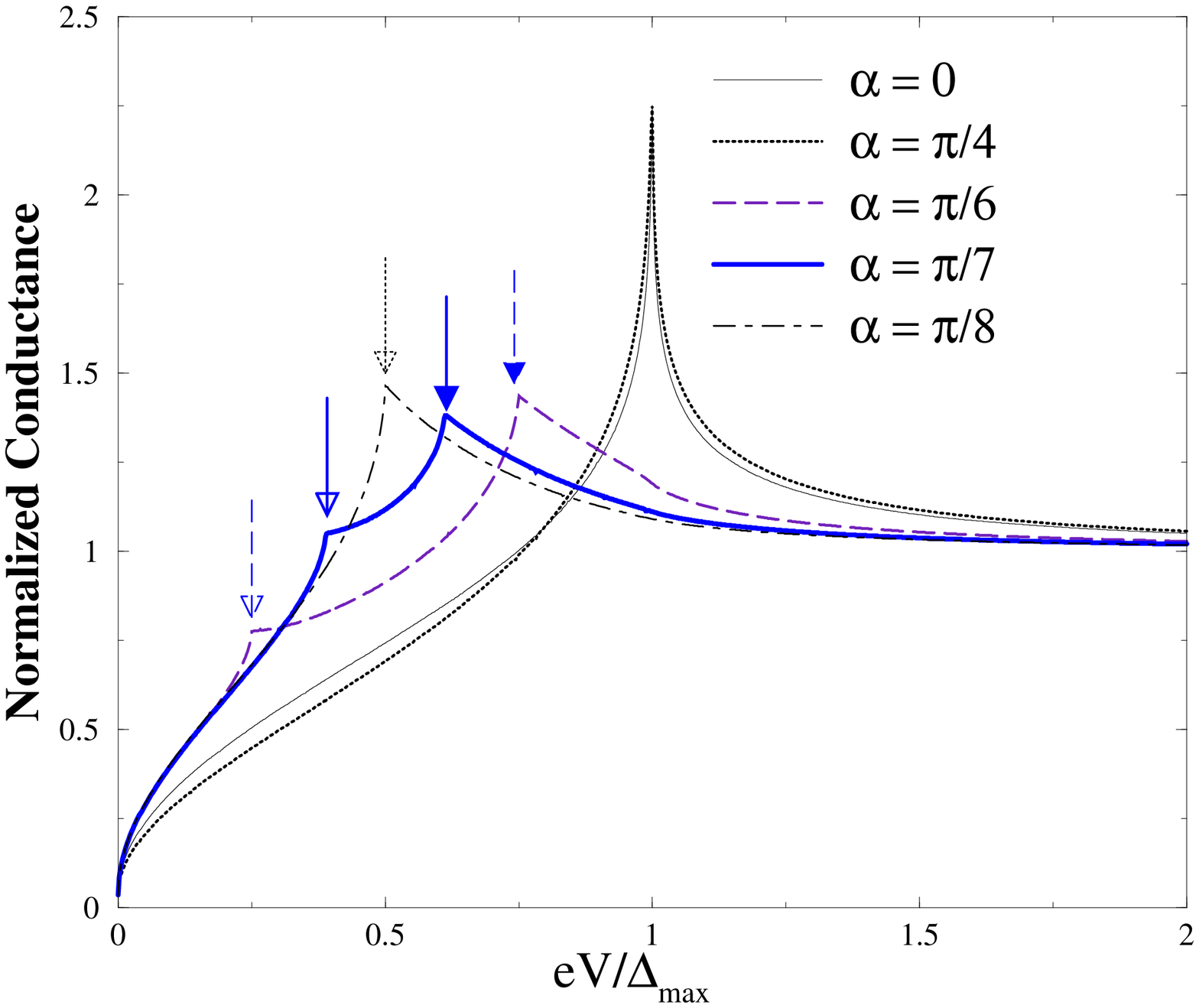}\\
{\small(b) $\Delta_s=0.9\Delta_g$}
\includegraphics[scale=0.4,angle=270]{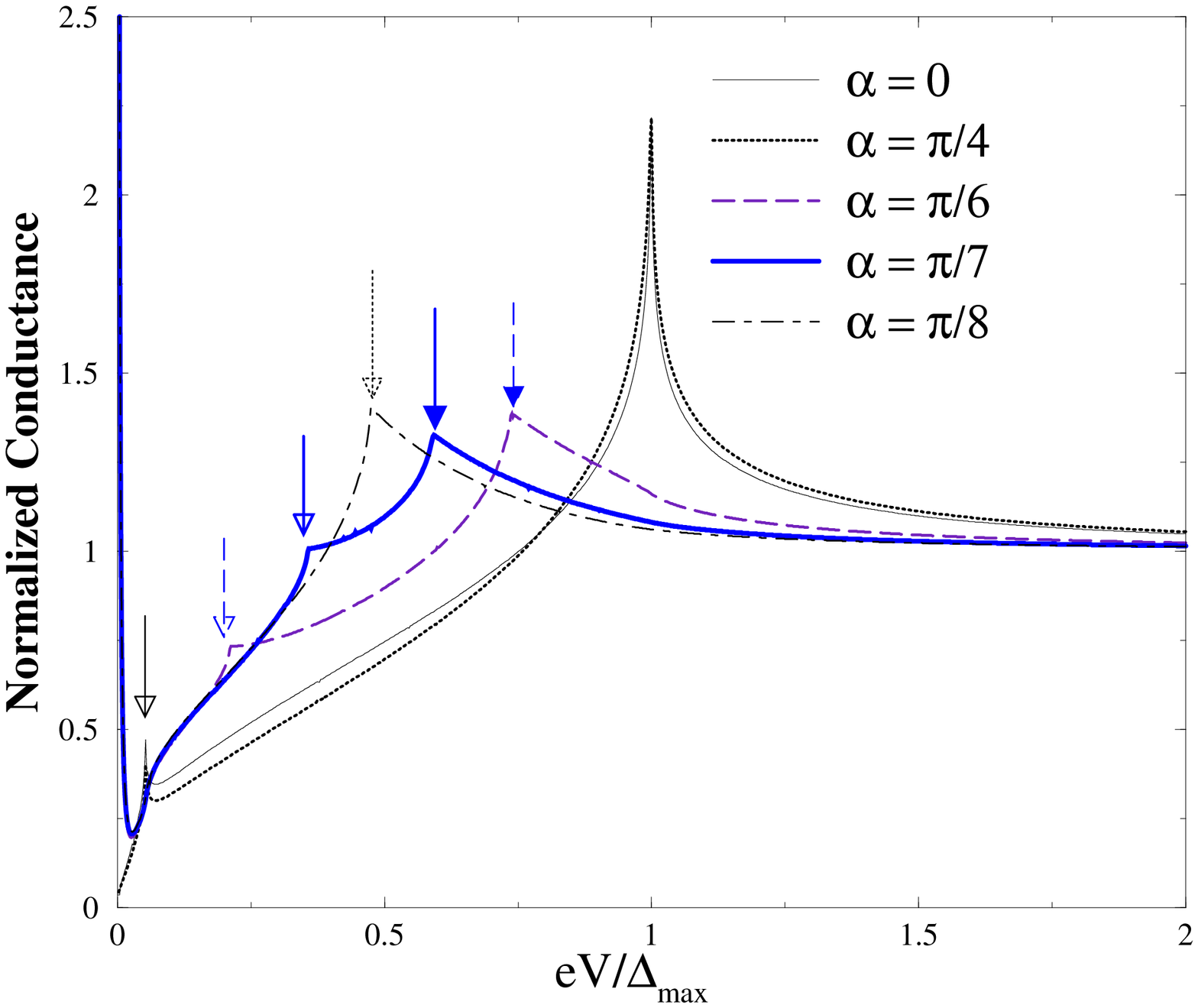}\\
{\small(c) $\Delta_s=1.1\Delta_g$}
\includegraphics[scale=0.4,angle=270]{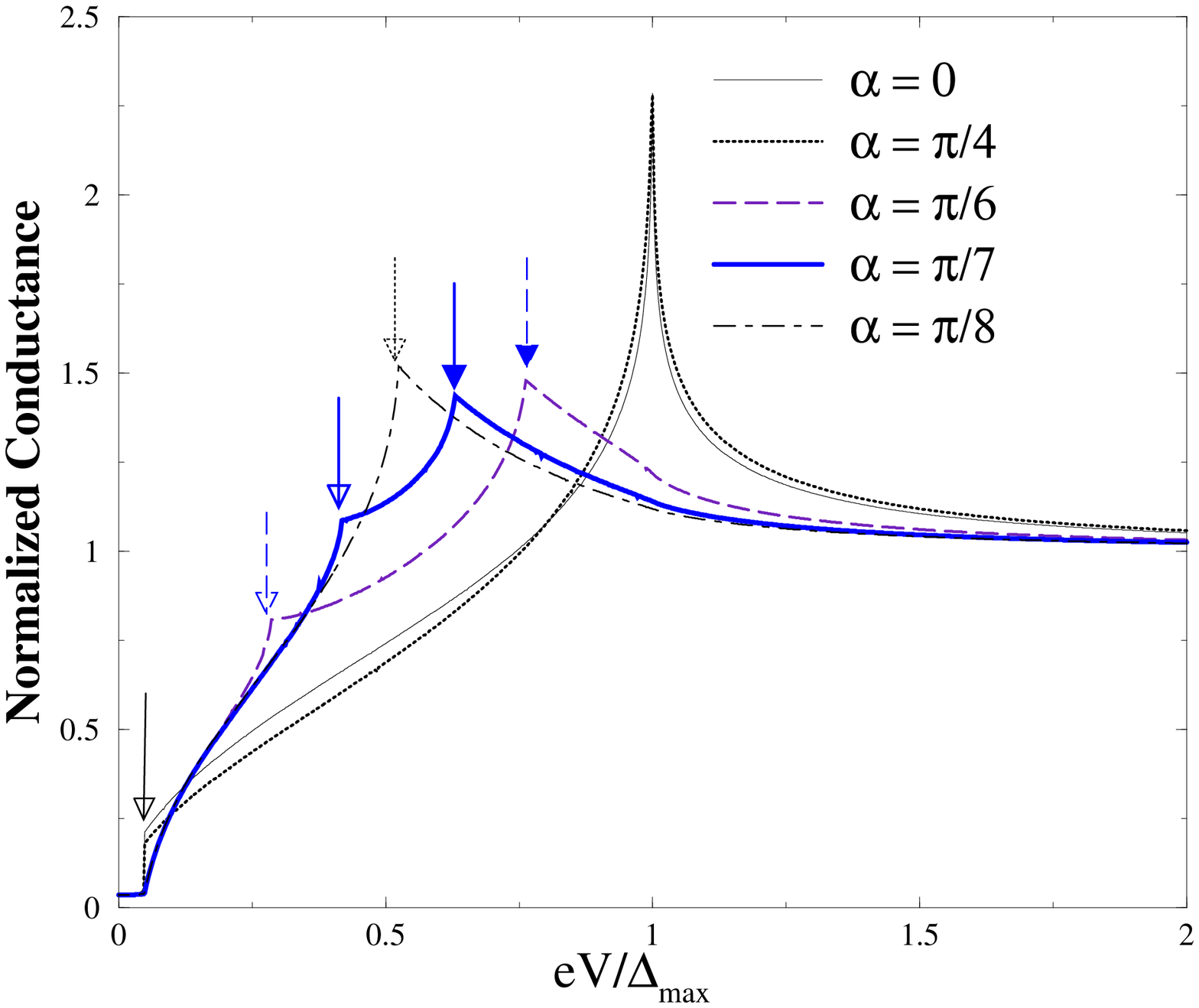}
\end{center}
\caption{\label{tun-abl-tunneling}$ab$ {\it plane tunnelling
junctions.} The plots of the normalized conductance spectra when
$Z=3$ for 3 interface orientations in 3 cases of the gap function
$\Delta_{k^\pm,1} = \Delta_s - \Delta_g\cos4(\phi_S\mp\alpha)$:
(a) $\Delta_s = \Delta_g$, (b) $\Delta_s = 0.9\Delta_g$, and (c)
$\Delta_s = 1.1\Delta_g$. The hollowed arrows indicate the
features occurring at voltage corresponding to
$\Delta_{k,1}(\theta_S=\pi/2,\phi_S=\pi/4,\alpha)$ and the filled
arrows mark the features at voltage corresponding to
$\Delta_{k,1}(\theta_S=\pi/2,\phi_S=0,\alpha)$. }
\end{figure}

\begin{figure}
\begin{center}
{\small(a) $\Delta_s=\Delta_g$}
\includegraphics[scale=0.4,angle=270]{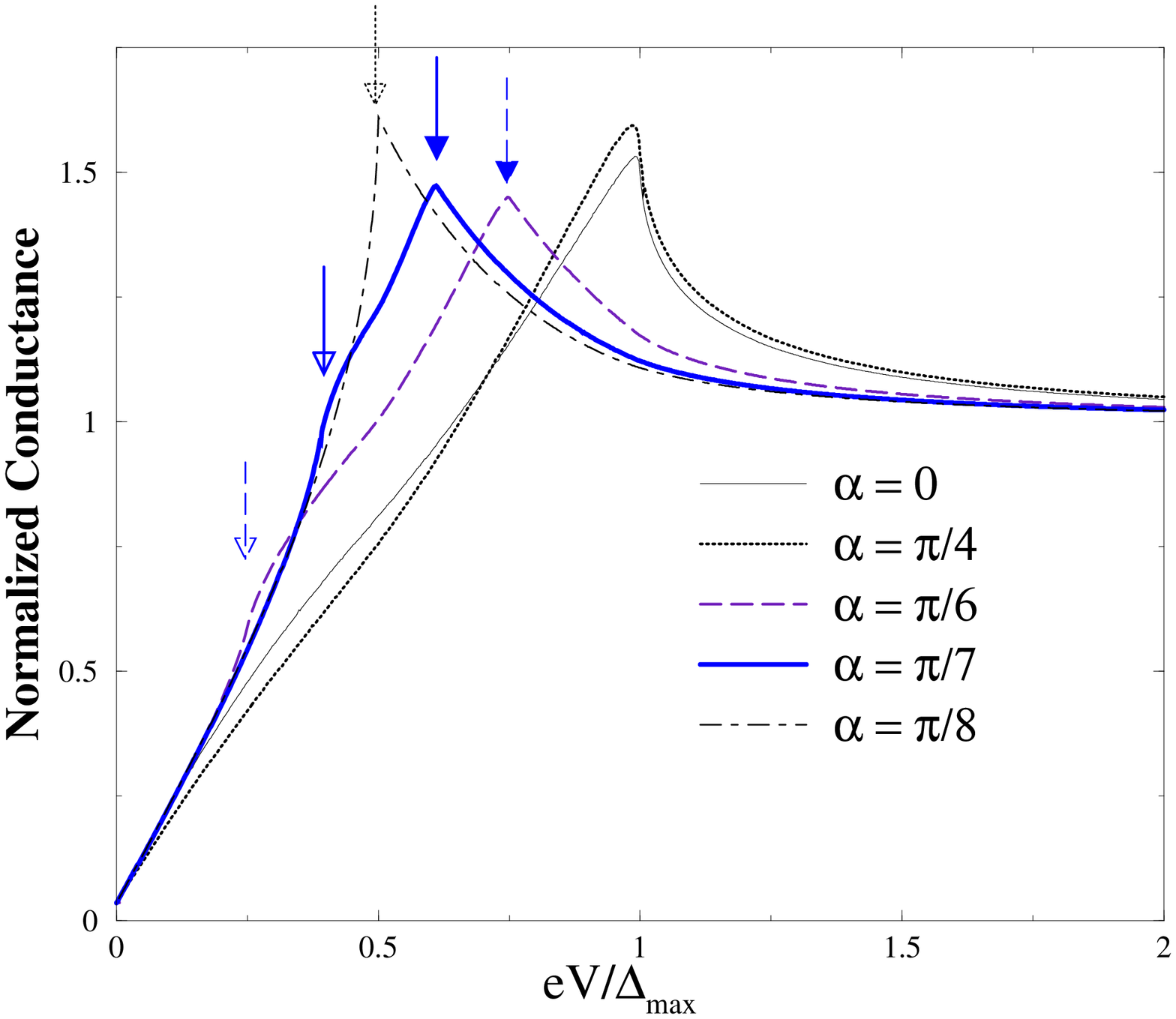}\\
{\small(b) $\Delta_s=0.9\Delta_g$}
\includegraphics[scale=0.4,angle=270]{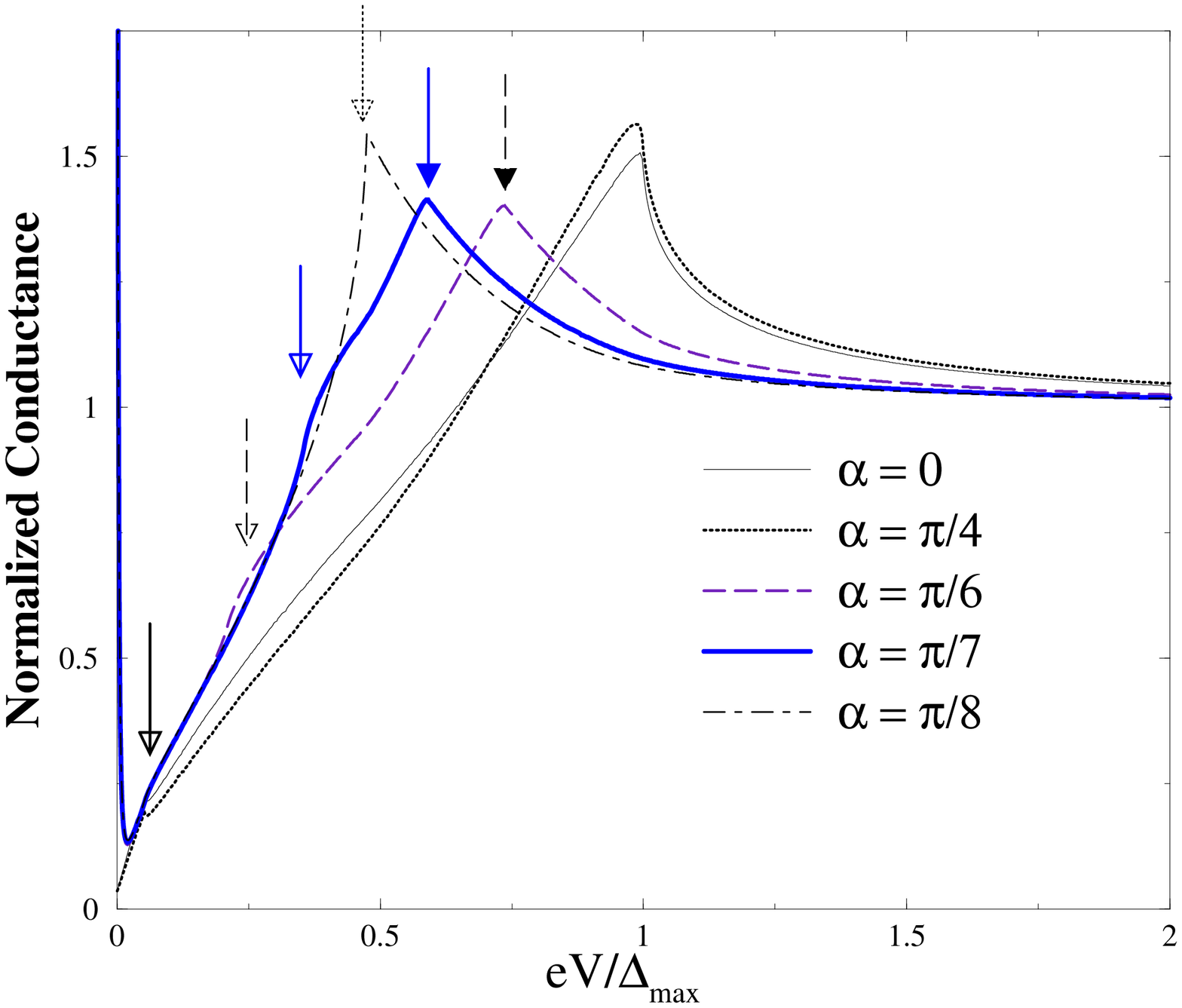}\\
{\small(c) $\Delta_s=1.1\Delta_g$}
\includegraphics[scale=0.4,angle=270]{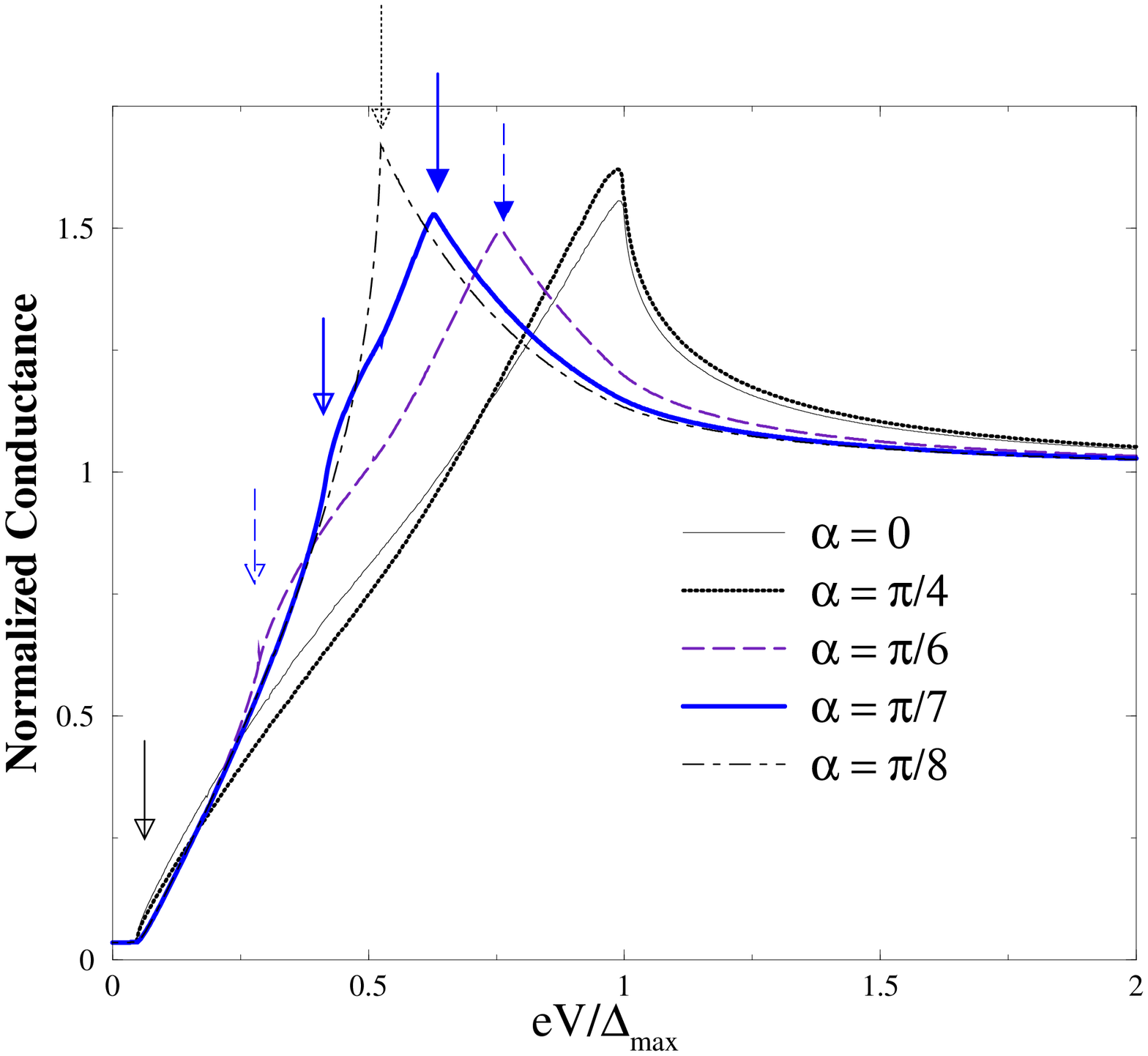}
\end{center}
\caption{\label{tun-abp-tunneling}$ab$ {\it plane tunnelling
junctions.} The plots of the normalized conductance spectra when
$Z=0$ for 3 interface orientations in 3 cases of the gap function
$\Delta_{k^\pm,2} = \Delta_s -
\Delta_g\sin^2\theta_S\cos4(\phi_S\mp\alpha)$: (a) $\Delta_s =
\Delta_g$, (b) $\Delta_s = 0.9\Delta_g$, and (c) $\Delta_s =
1.1\Delta_g$. The hollowed arrows indicate the features at voltage
corresponding to
$\Delta_{k,2}(\theta_S=\pi/2,\phi_S=\pi/4,\alpha)$ and the filled
arrows mark the features at voltage corresponding to
$\Delta_{k,2}(\theta_S=\pi/2,\phi_S=0,\alpha)$. }
\end{figure}

In the tunnelling limit (large $Z$), the shape of the conductance
spectrum changes with the interface orientation. First, consider
the case where $\Delta_s = \Delta_g$, there are two distinct peaks
which are more pronounced for the superconductor with the gap
function $\Delta_{k,1}$ than for the superconductor with the gap
function $\Delta_{k,2}$ (compare figure~\ref{tun-abl-tunneling}
(a) and figure~\ref{tun-abp-tunneling} (a)).
The two peaks occur at voltages corresponding to
$\Delta_{k,i}(\theta_S=\pi/2,\phi_S=0,\alpha)$ (marked by the
filled arrows) and
$\Delta_{k,i}(\theta_S=\pi/2,\phi_S=\pi/4,\alpha)$ (marked by the
hollowed arrows), where $i$ is either 1 or 2. Notice that when
$\alpha=\pi/8$, there is only one peak due to the fact that
$\Delta_{k,i}(\theta_S=\pi/2,\phi_S=0,\alpha=\pi/8) =
\Delta_{k,i}(\theta_S=\pi/2,\phi_S=\pi/4,\alpha=\pi/8)$.

In the case of $\Delta_s=0.9\Delta_g$, $\Delta_{k,i}$ can be both
positive and negative. Consequently, for the orientations with
$\alpha\neq 0,\pi/4$, in addition to the two peaks at the
positions like those in the case of $\Delta_s=\Delta_g$, there
exists a peak at zero voltage (see figure~\ref{tun-abl-tunneling}
(b) and \ref{tun-abp-tunneling} (b)). The existence of this
zero-bias conductance peak, which also occurs in the $d$-wave
case, is a signature of the presence of a sign change of the gap
function~\cite{hu94,kashi95}.

When $\Delta_s=1.1\Delta_g$, there is a finite gap minimum. The
conductance is very small for voltages less than the gap minimum,
and the two peaks at
$\Delta_{k,i}(\theta_S=\pi/2,\phi_S=0,\alpha)$ and
$\Delta_{k,i}(\theta_S=\pi/2,\phi_S=\pi/4,\alpha)$ are still
present.
\begin{figure}
\begin{center}
\includegraphics[scale=0.7]{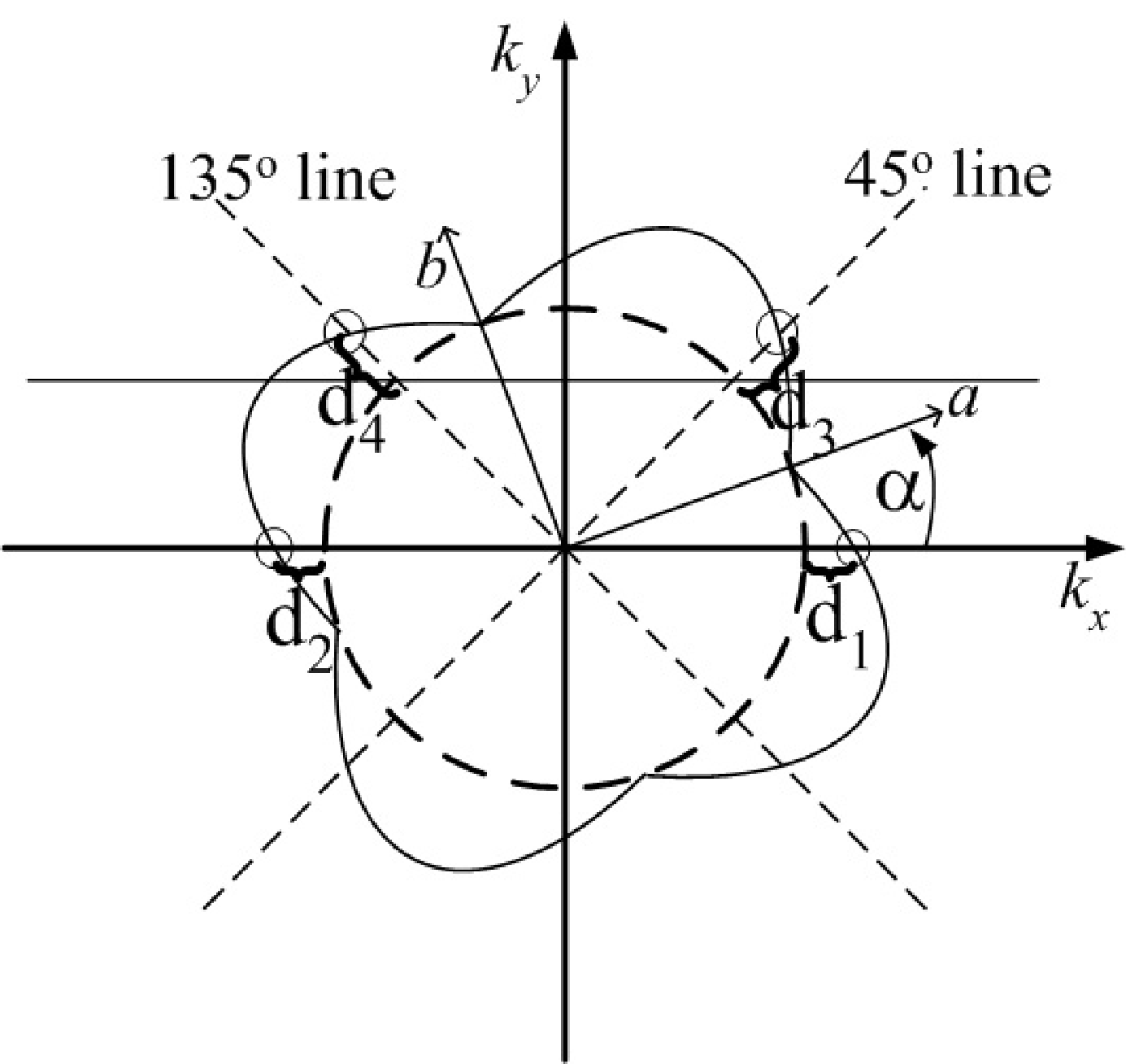}
\end{center}
\caption{\label{gap-plot}This picture shows a sketch of the gap
function in the momentum space, when $\alpha\neq 0,\pi/4$. The
dashed circle is the contour of the Fermi surface projected on the
$k_xk_y$ plane. The solid curve represents the gap function. The
two dashed lines are the $\phi_S=\pi/4$ and $\phi_S=3\pi/4$ lines.
The solid horizontal line above $k_y=0$ is the line of constant
$k_y$ at $\phi_S=\pi/4$ on the Fermi surface. This line cute the
Fermi surface at 2 points on the plane. Note that $d_1=d_2$ (the
magnitudes of the gap of the two excitations when $\phi_S=0$) and
$d_3=d_4$ (the magnitudes of the gap the two excitations when
$\phi_S=\pi/4$).}
\end{figure}

In summary, the conductance spectrum in every case where $\alpha
\neq 0, \pi/4$ contains two peaks at
$\Delta_{k,i}(\theta_S=\pi/2,\phi_S=0,\alpha)$ and
$\Delta_{k,i}(\theta_S=\pi/2,\phi_S=\pi/4,\alpha)$. These two
angles, $\phi_S = 0,\pi/4$, are special because the magnitudes of
the gap function of the two transmitted superconducting
excitations with the $k_y$ corresponding to these angles are the
same, i.e., $\Delta_{k^+} = \Delta_{-k^-}$, as shown pictorially
in figure~\ref{gap-plot}. The feature at
$\Delta_{k,i}(\theta_S=\pi/2,\phi_S=0,\alpha)$ has also been shown
to occur in a $d$-wave superconductor in a continuous
model~\cite{pairor02}. However, the feature at
$\Delta_{k,i}(\theta_S=\pi/2,\phi_S=\pi/4,\alpha)$ is unique to an
anisotropic $s$-wave superconductor in this model. In a d-wave
superconductor the values of
$\Delta_{k^+}(\theta_S=\pi/2,\phi_S=\pi/4,\alpha)$ and
$\Delta_{-k^-}(\theta_S=\pi/2,\phi_S=\pi/4,\alpha)$ always have
opposite signs and thus excitations having these momenta
contribute to the zero energy surface bound state~\cite{hu94}
instead of giving rise to such a feature. We note that the two
peaks at $\Delta_{k,i}(\theta_S=\pi/2,\phi_S=0,\alpha)$ and
$\Delta_{k,i}(\theta_S=\pi/2,\phi_S=\pi/4,\alpha)$ are more
prominent for the superconductor with the gap function
$\Delta_{k,1}$ than for the superconductor with the gap function
$\Delta_{k,2}$. Given only this quantitative distinction between
the $ab$ plane tunnelling results for the two forms of the gap
considered, it would likely be difficult to determine from real
tunnelling data which form is present. The $ab$ plane tunnelling
spectrum would enable a determination of the magnitude of the gap
function in an arbitrary direction in the plane for either of the
two forms of the gap.

\subsection{$c$-axis tunnelling}
\begin{figure}
\begin{center}
{\small(a) $Z=0$}
\includegraphics[scale=0.4,angle=270]{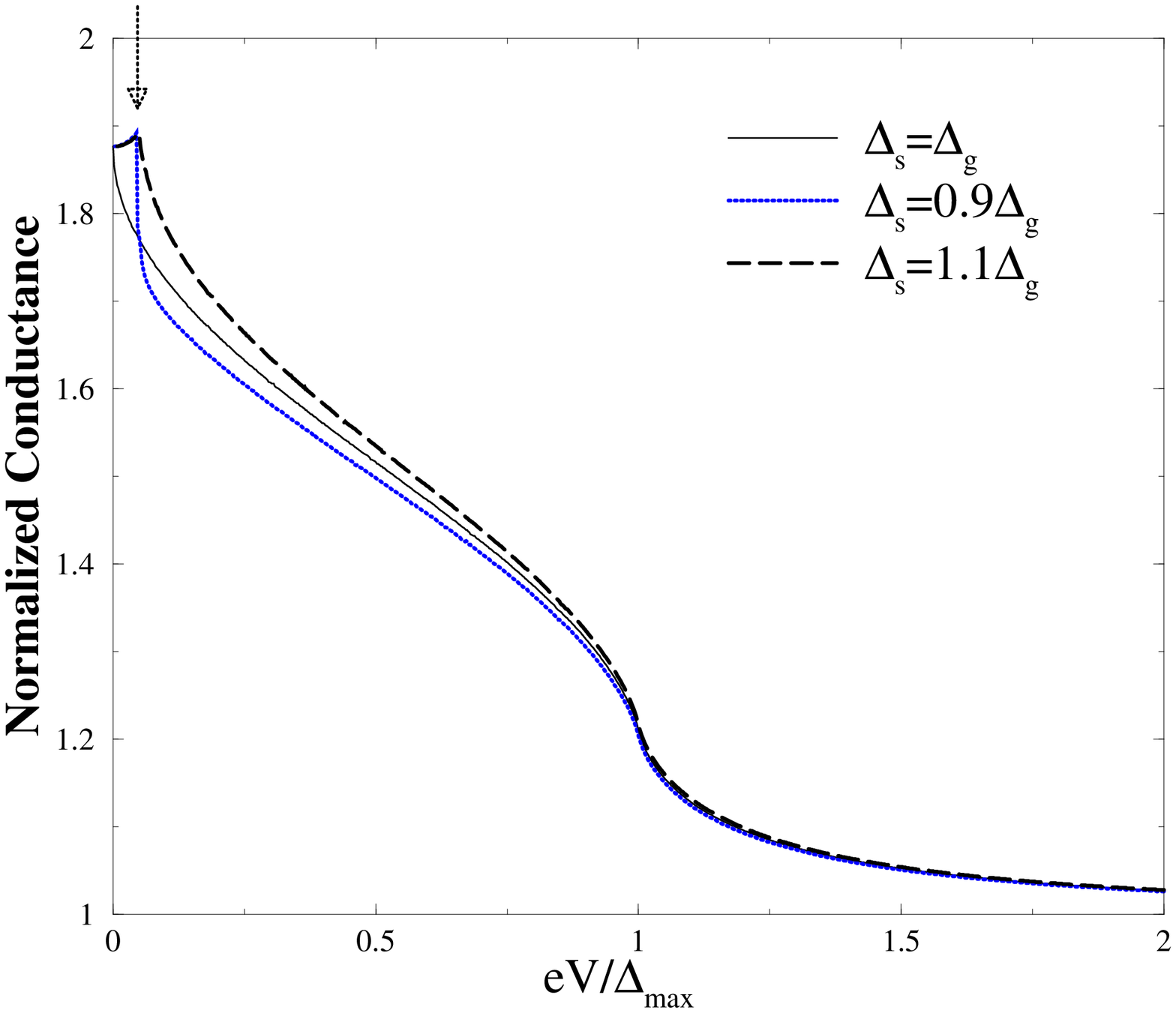}\\
{\small(b) $Z=3$}
\includegraphics[scale=0.4,angle=270]{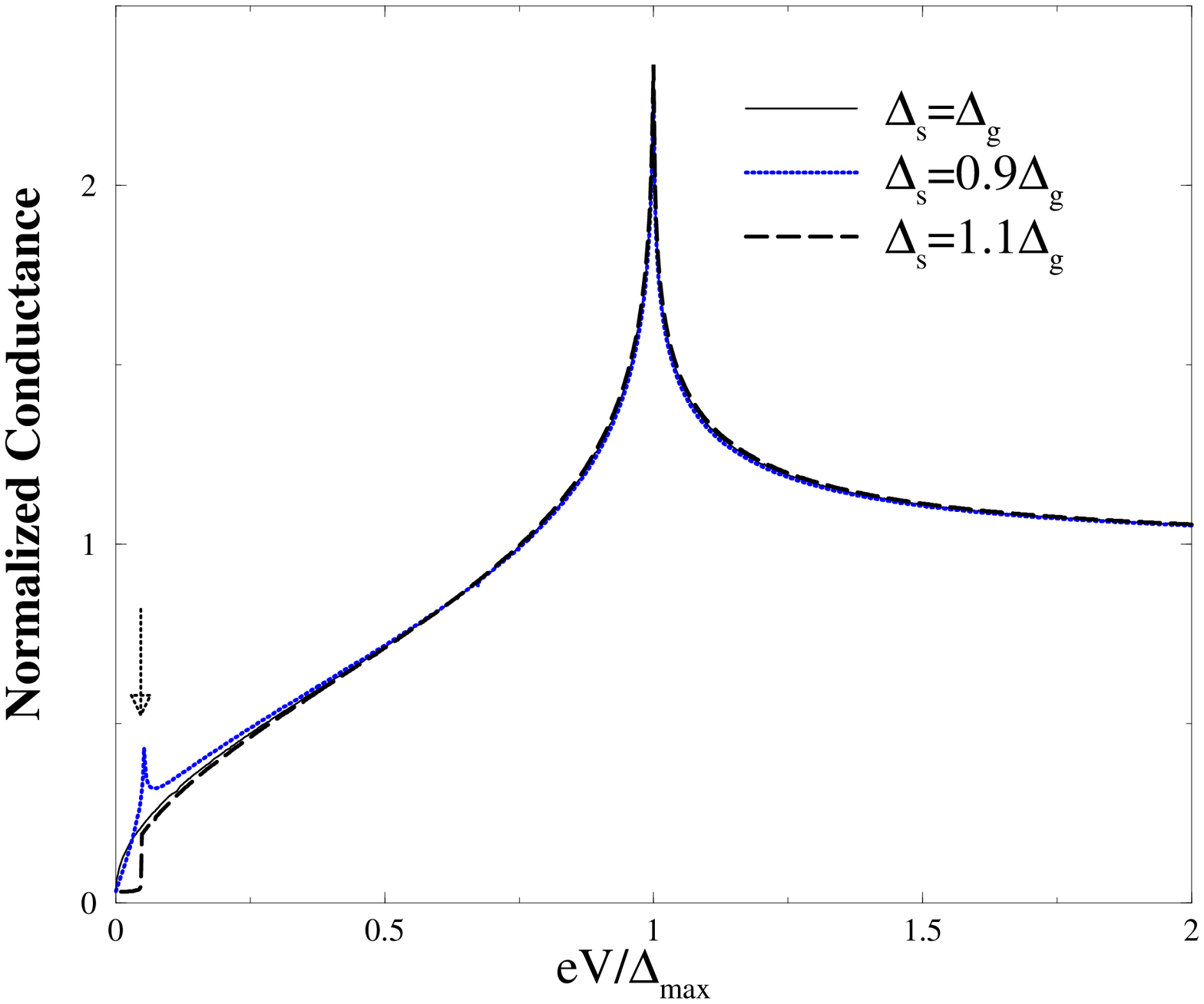}
\end{center}
\caption{\label{cl-tunneling}{\bf $c$}{\it-axis tunnelling
junctions.} The plots of the normalized conductance spectra for
(a) $Z=0$ and (b) $Z=3$ in case of the gap $\Delta_{k,1} =
\Delta_s - \Delta_g\cos4(\phi_S\mp\alpha)$. The arrows indicate a
feature occurring at $|\Delta_s-\Delta_g|$.}
\end{figure}

\begin{figure}
\begin{center}
{\small(a) $Z=0$}
\includegraphics[scale=0.4,angle=270]{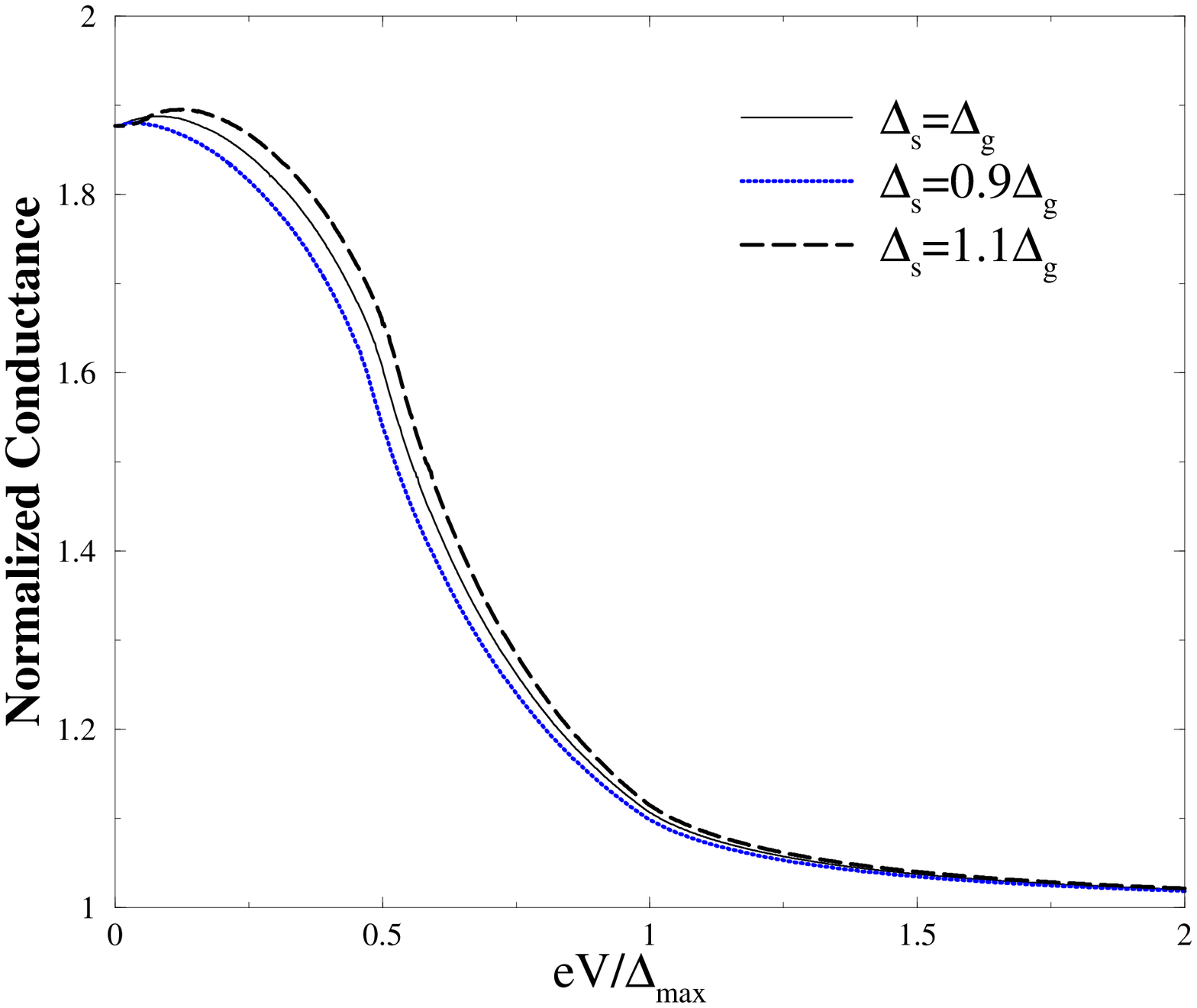}\\
{\small(b) $Z=3$}
\includegraphics[scale=0.4,angle=270]{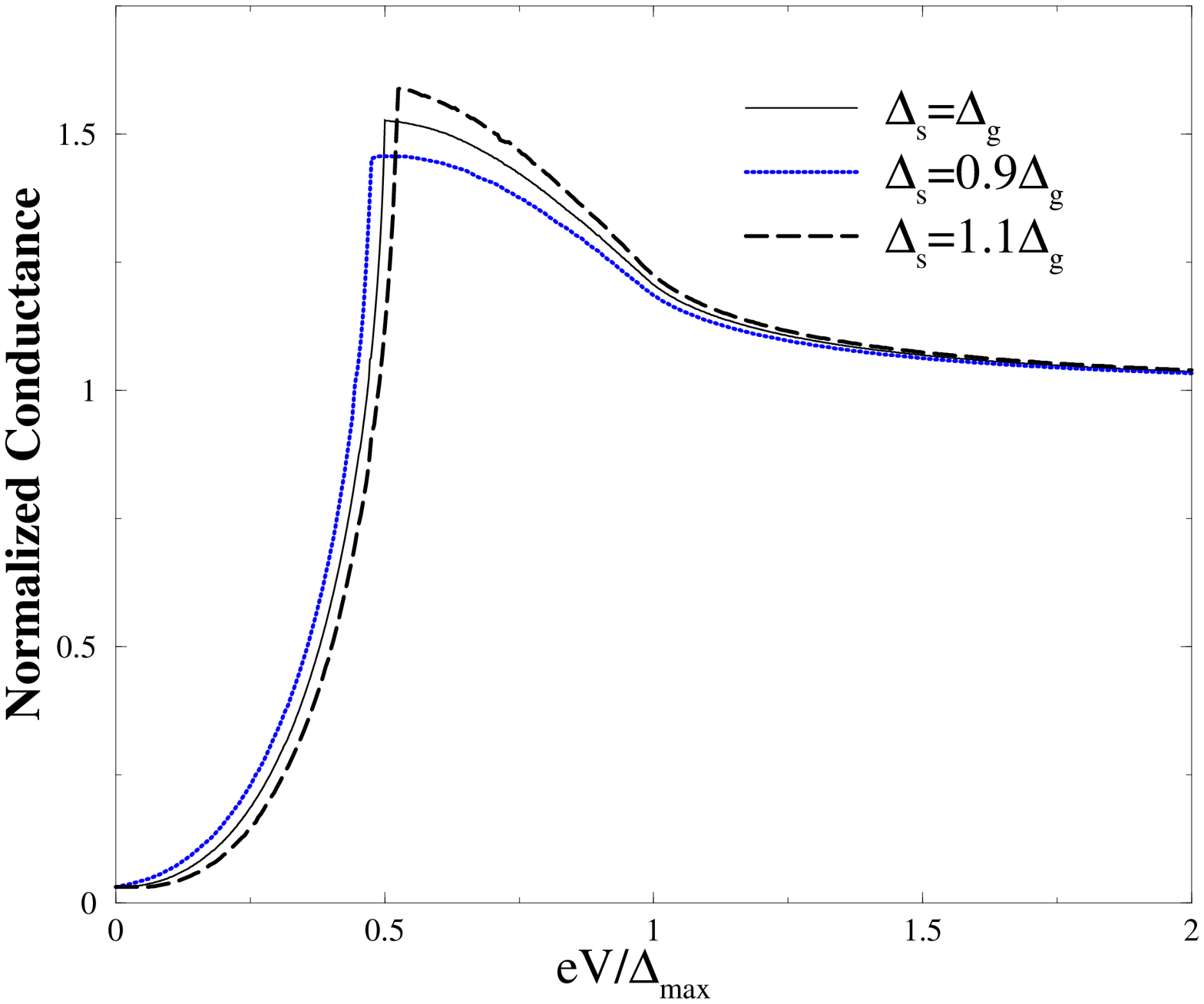}
\end{center}
\caption{\label{cp-tunneling}{\bf $c$}{\it-axis tunnelling
junctions.} The plots of the normalized conductance spectra for
(a) $Z=0$ and (b) $Z=3$ in case of the gap $\Delta_{k,2} =
\Delta_s - \Delta_g\sin^2\theta_S\cos4(\phi_S\mp\alpha)$.}
\end{figure}

Because in the continuous model the Fermi surface of a tetragonal
crystal is invariant under rotation around the $c$ axis of the
crystal, the $c$-axis tunnelling spectroscopy is independent of
the rotation around the $c$ axis. In the Andreev limit, when $
|\Delta_s-\Delta_g|<$ eV $<\Delta_{max}$, the conductance curve of
the superconductor with the gap function $\Delta_{k,1}$ is upward
and decreasing smoothly (see figure~\ref{cl-tunneling} (a)). On
the contrary, the conductance curve of the superconductor with the
gap function $\Delta_{k,2}$ is downward when
$|\Delta_s-\Delta_g|<$ eV $< \Delta_{k,2}(\theta_S=0)$ (see
figure~\ref{cp-tunneling} (a)).


In the tunnelling limit, the spectrum of the superconductor with
the gap function $\Delta_{k,1}$ contains a peak at the gap maximum
and a feature at $|\Delta_s-\Delta_g|$ (see
figure~\ref{cl-tunneling} (b)). The spectrum of the superconductor
with the gap function $\Delta_{k,2}$ contains a feature at the gap
maximum, and a sharp peak at $\Delta_{k,2}(\theta_S=0)$ (see
figure~\ref{cp-tunneling} (b)). The occurrence of the sharp peak
at different positions in the conductance spectrum makes it
possible to distinguish  the tunnelling spectrum of the line-node
gap from the point-node gap form using $c$-axis tunnelling data.

\section{Conclusions}

We have studied the $c$-axis and $ab$ plane tunnelling
spectroscopy of $s+g$-wave superconductors. The observation of the
predicted features in tunnelling measurements made for various
junction orientations would, in principle, provide a way to study
the detailed momentum dependence of an $s$+$g$-wave
superconducting gap. In borocarbides, the presence of an
$s+g$-wave gap function having either line or point nodes has been
suggested. Directional tunnelling spectroscopy would help to
determine whether either form of the superconducting gap is
correct.

We can distinguish the $c$-axis tunnelling spectra of a
superconductor with line nodes from the spectra of a
superconductor with point nodes. In the tunnelling limit, the
$c$-axis tunnelling conductance spectrum of a line-node
superconductor contains a sharp peak at the gap maximum, whereas
in the spectrum of a point-node superconductor a sharp peak occurs
at the value of the gap function along the $c$ axis.

The $ab$ plane tunnelling spectra may be used to map out the
magnitude of the gap function in the plane. The conductance
spectrum of $ab$ plane tunnelling junction is strongly dependent
on the junction orientation. Two features in the spectrum appear
at energies equal to the magnitude of the superconducting gap in
two momentum directions: one is the direction parallel to the
interface normal and the other is the direction making a $\pi/4$
angle with the interface normal. It is worth noting that these
features are more prominent in the spectrum of the superconductor
with line nodes than in the spectrum of the superconductor with
point nodes, although this subtle distinction may not be
observable especially at finite temperatures. \vspace{1pc}

\ack We would like to thank M. B. Walker for valuable discussions
and comments. Also, PP would like to thank Thai Research Fund
(TRF, grant no. TRG4580057) for financial support.
\\

\end{document}